\documentclass[twocolumn]{aastex63}%,trackchanges
\usepackage{amsmath,color,textcomp,url,graphicx}

\usepackage{subfigure}
\makeatletter

\renewcommand{\@thesubfigure}{(\alph{subfigure})\hskip\subfiglabelskip}
\renewcommand{\@@thesubfigure}{(\alph{subfigure})}
\makeatother

\newcommand{\kms}{\hbox{km\,s$^{-1}$}}

\submitjournal{ApJL}

\shorttitle{A Number of Nearby Moving Groups May Be Fragments of Dissolving Open Clusters}
\shortauthors{Gagn\'e et al.}

\begin{document}

\title{A NUMBER OF NEARBY MOVING GROUPS MAY BE FRAGMENTS OF DISSOLVING OPEN CLUSTERS}

\author[0000-0002-2592-9612]{Jonathan Gagn\'e}
\affiliation{Plan\'etarium Rio Tinto Alcan, Espace pour la Vie, 4801 av. Pierre-de Coubertin, Montr\'eal, Qu\'ebec, Canada}
\affiliation{Institute for Research on Exoplanets, Universit\'e de Montr\'eal, D\'epartement de Physique, C.P.~6128 Succ. Centre-ville, Montr\'eal, QC H3C~3J7, Canada}
\email{gagne@astro.umontreal.ca}
\author[0000-0001-6251-0573]{Jacqueline K. Faherty}
\affiliation{Department of Astrophysics, American Museum of Natural History, Central Park West at 79th St., New York, NY 10024, USA}
\author[0000-0001-7171-5538]{Leslie Moranta}
\affiliation{Plan\'etarium Rio Tinto Alcan, Espace pour la Vie, 4801 av. Pierre-de Coubertin, Montr\'eal, Qu\'ebec, Canada}
\affiliation{Institute for Research on Exoplanets, Universit\'e de Montr\'eal, D\'epartement de Physique, C.P.~6128 Succ. Centre-ville, Montr\'eal, QC H3C~3J7, Canada}
\author[0000-0001-9482-7794]{Mark Popinchalk}
\affil{Department of Astrophysics, American Museum of Natural History, Central Park West at 79th Street, NY 10024, USA }
\affiliation{Physics, The Graduate Center, City University of New York, New York, NY 10016, USA}
\affiliation{Department of Physics and Astronomy, Hunter College, City University of New York, 695 Park Avenue, New York, NY 10065, USA}
\email{popinchalkmark@gmail.com}

\begin{abstract}

We propose that fourteen co-moving groups of stars uncovered by \cite{2019AJ....158..122K} may be related to known nearby moving groups and bridge those and nearby open clusters with similar ages and space velocities. This indicates that known nearby moving groups may be spatially much more extended than previously though, and some of them might be parts of tidal tails around the cores of known open clusters, reminiscent of those recently found around the Hyades and a handful of other nearby clusters. For example, we find that both the nearby Carina and Columba associations may be linked to Theia~208 from \cite{2019AJ....158..122K} and together form parts of a large tidal tail around the Platais~8 open cluster. The AB~Doradus moving group and Theia~301 may form a trailing tidal tail behind the Pleiades open cluster, with hints of a possible leading tidal tail in Theia~369. We similarly find that IC~2391 and its tidal tails identified by \cite{2021AA...645A..84M} may be extended by the nearby Argus association and are possibly further extended by Theia~115. The nearby Octans and Octans-Near associations, as well as Theia~94 and 95, may form a large tidal tail leading the poorly studied Platais~5 open cluster candidate. While a preliminary analysis of Gaia color-magnitude sequences hint that these structures are plausibly related, more observational evidence is still required to corroborate their consistent ages and space velocities. These observations may change our current understanding of nearby moving groups and the different pathways through which they can form. While some moving groups may have formed loosely in extended star-formation events with rich spatial structure, others may in fact correspond to the tidal tails of nearby open clusters.

\end{abstract}


\section{INTRODUCTION}\label{sec:intro}

Moving groups were historically discovered as sparse ensembles of stars that caused distinct clumps in the $UVW$ space velocity distribution of young stars near the Sun \citep{1973RPPh...36..625E}. As decades passed and the field of stellar kinematics progressed, several moving groups were discovered and characterized \citep{2004ARAA..42..685Z}, with spatial distributions restricted to about $\sim$\,100\,pc from the Sun, due to the limited accuracy and sensitivity of past surveys. Some of these co-moving groups of young stars were named `young associations' rather than moving groups, such as the Tucana-Horologium, Carina and Columba associations \citep{2000AJ....120.1410T,2008hsf2.book..757T} despite their being similarly sparse and in close proximity to the Sun. Individual sparse young associations near the Sun were generally thought to have formed together from a distinct molecular cloud, although some of them were recognized as being plausibly related to more massive nearby open clusters due to their similar ages and space velocities. Such examples included a potential relation between the AB~Doradus moving group \citep{2004ApJ...613L..65Z} and the Pleiades open cluster \citep{2005ApJ...628L..69L}, as well as between the Argus association and the IC~2391 open cluster \citep{2000MNRAS.317..289M}. However, the details of how these associations may be related were never clearly elucidated.

The brightest and most massive members of nearby young associations and moving groups were the first to be discovered and characterized, given their detections in X-ray surveys such as ROSAT \citep{2016AA...588A.103B} and astrometric catalogs such as Hipparcos \citep{1997AA...323L..49P}. Follow-up studies used prior knowledge of these distributions of these nearby stars in $XYZ$ Galactic coordinates and $UVW$ space velocities to identify the lower-mass members which dominate the mass function. Such studies typically worked directly in sky coordinates and proper motion space \citep{2005ApJ...634.1385M} or used Bayesian model selection methods to capture the missing low-mass members \citep{2013ApJ...762...88M,2018ApJ...856...23G}. The recent data releases of the Gaia mission \citep{2016AA...595A...1G} allowed these studies to uncover a large number of missing M dwarfs \citep{2018ApJ...862..138G}, however, none of the aforementioned methods were able to detect spatial extensions of moving groups, by construction.

The Gaia mission provided the scientific community with a vast number of high-precision stellar kinematics, which allowed large-scale identification of co-moving stars for the first time. \cite{2017AJ....153..257O} pioneered such searches by identifying thousands of co-moving pairs of stars with Gaia~DR1 \citep{2016AA...595A...2G} and serendipitously uncovered known and new young associations. More recently, \citet[KC19 hereafter]{2019AJ....158..122K} performed a search based on Gaia~DR2 \citep{GaiaCollaboration:2018io} designed specifically to uncover extended and low-density streams of co-moving, co-eval stars. This allowed them to identify 1640 co-moving streams of stars with distances 80--1000\,pc (with the exception of the dense Hyades at $\sim$\,50\,pc). Similar to the work of \cite{2017AJ....153..257O}, the method of KC19 relied on clustering algorithms that work directly in observable space (sky position, proper motion, parallax and radial velocity) because even after Gaia~DR2, we still lack radial velocity measurements for the vast majority of nearby stars. This prevented these searches from uncovering significant clusters of co-moving stars closer than about 70\,pc, because the direct observables become widely distributed and highly correlated with sky position. Clustering algorithms therefore either fail to uncover them without a large number of false-positives, or break up each nearby association into many smaller parts \citep{2018ApJ...863...91F}.

In parallel to these developments, Gaia-enabled searches in velocity space resulted in the discovery of extended tidal tails associated with the Hyades and Coma~Berenices clusters \citep{2019AA...621L...2R,2019ApJ...877...12T}, and more recently to 9 of the nearest open clusters \citep{2021AA...645A..84M}. Candidate tidal tails associated with the Ursa~Major core of 10 massive stars were also recently identified by \cite{2020RNAAS...4...92G}. Gaia also allowed the study of young star-forming regions with unprecedented detail, which allowed recent work to describe the complex structure and large spatial extent of Taurus \citep{2021arXiv210513370K,Liu:2021ep}, Orion, Perseus and Sco-Cen \citep{Kerr:2021uz}. These discoveries are enriching our current understanding of how complex the spatial structures of both open clusters and loose star-forming regions can be, and hint that parts of them may be related to what we have studied under the name of moving groups for a few decades.

In this Letter, we propose that several known moving groups and young associations near the Sun may be related to extended co-moving structures uncovered by KC19, as well as known and slighly more distant open clusters and young associations, based on their similar space velocities and ages. We posit that dedicated algorithms and follow-up observations will be needed to test this hypothesis, by filling out current gaps within 70\,pc of the Sun, reducing sample contamination, and elucidating the exact spatial extent of known young moving groups near the Sun. We suggest that the biggest challenge in this follow-up is the fact that the properties of nearby sparse associations are spread out and correlated with sky position because of purely geometric effects. A successful verification that a majority of young moving groups are indeed related to more distant open clusters and extended young associations would shift our interpretation of how they form and evolve, and would open the door to assembling much larger ensembles of co-eval stars that are valuable laboratories to study the fundamental properties and evolutions of stars, brown dwarfs and exoplanets.

\section{SAMPLE AND METHOD}\label{sec:sample_method}

We investigated the Theia groups identified by KC19 to determine whether some of them may be associated either with known nearby young moving groups, associations and open clusters. In order to do so, we pre-selected all Theia groups with a Gaia~DR2 $G - G_{\rm RP}$ color versus absolute $G$ magnitude diagram consistent with that of a currently known group of stars, and we then visualized all of the resulting groups in $XYZ$ Galactic coordinates as well as $UVW$ space velocities using the Partiview 3-dimensional visualization software \citep{2003IAUS..208..343L}. We have elected to use Gaia~DR2 data in this work to remain consistent with the KC19 groups without having to re-define or re-characterize them; note that a proper comparison of Gaia color-magnitude diagrams in particular requires to account for the different photometric bandpasses between different Gaia data releases.

Every Theia group already comes with an age estimate in KC19, which were determined using the PARSEC model isochrones \citep{2017ApJ...835...77M}. However, during this investigation we realized that this method seemed overly dependent on outlier data points and did not produce ages consistent with those of known young associations with similar color-magnitude sequences, likely because of systematics that evolutionary models are known to produce when using them to calculate isochrone ages \citep{2015MNRAS.454..593B}. We have therefore revised the most likely ages of every Theia group with an empirical approach. In a first step, we built five Gaia~DR2 $G$ versus $G - G_{\rm RP}$ empirical color-magnitude sequences using groups of well-known ages. We used members of the Coma Berenices open cluster \citep{2006MNRAS.365..447C} as a $\sim$\,600\,Myr reference (\citealt{2014AA...566A.132S} find an age of $560_{-80}^{+100}$\,Myr), and members of the Pleiades open cluster ($112 \pm 5$\,Myr; \citealt{2015ApJ...813..108D}) as a $\sim$\,110\,Myr reference. We combined members of the $\sim$\,45\,Myr-old coeval Carina, Columba and Tucana-Horologium associations, and the members of the coeval $\sim$\,23\,Myr-old $\beta$~Pictoris moving group \citep{2001ApJ...562L..87Z} and the $22_{-3}^{+4}$\,Myr-old 32~Ori association \citep{2007IAUS..237..442M} as two additional reference sequences. In addition to those, we included a fifth reference sequence built from members of the 10--15\,Myr-old Sco-Cen OB association \citep{1946PGro...52....1B}. We used the lists of bona fide members from \cite{2018ApJ...856...23G} to build these respective samples of stars.

\begin{figure}
	\centering
	\includegraphics[width=0.48\textwidth]{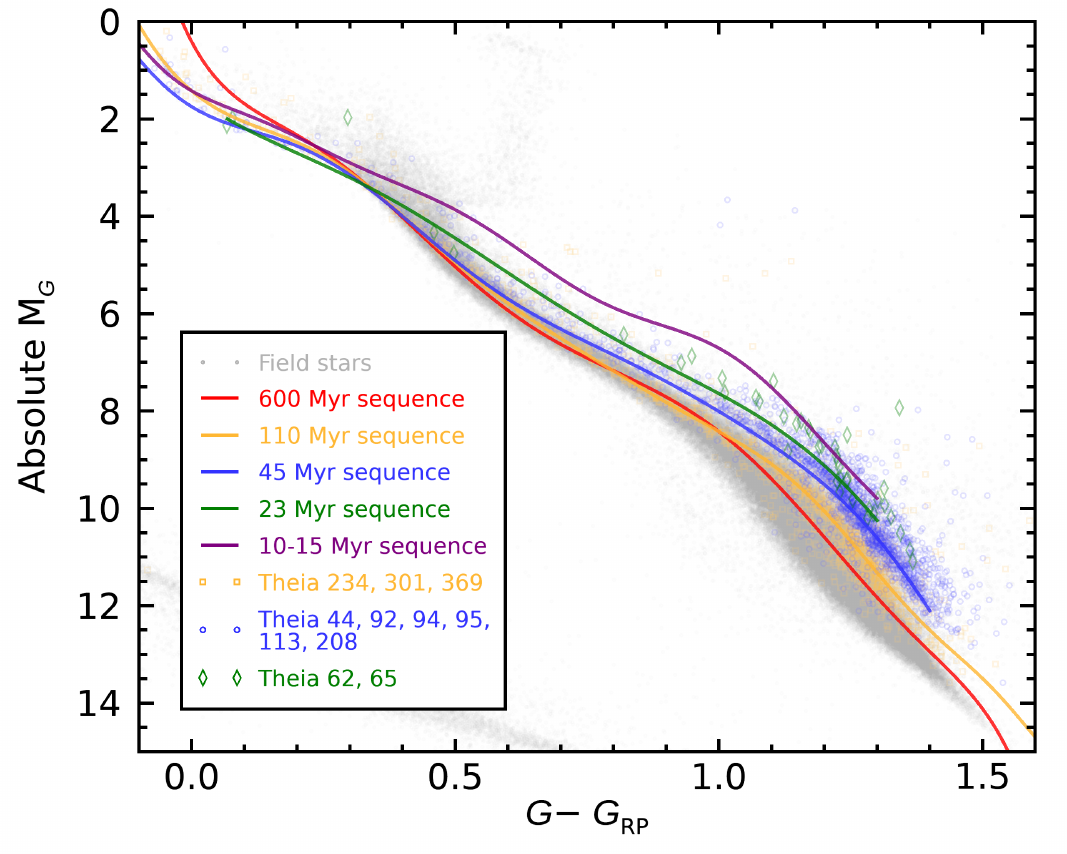}
	\label{fig:comd}
	\caption{Empirical sequences of known co-eval populations of stars in a Gaia~DR2 color-magnitude diagrams, as constructed and described in Section~\ref{sec:sample_method}. Members of relevant KC19 Theia groups discussed in Section~\ref{sec:results} are shown with pale symbols with colors corresponding to their best-matching empirical sequences.}
\end{figure}

We fit a polynomial sequence to each of these samples by first binning the color-magnitude sequences in 0.05\,mag-wide bins in Gaia~DR2 $G - G_{\rm RP}$ colors. The average and standard deviations of the absolute Gaia~DR2 $G$-band magnitudes of all stars falling in a given bin were therefore calculated, and subsequently fit with a 8- to 15-order polynomial, depending mostly on the range of colors and the total number of data points in each sample (provided online as data behind figure). Individual measurement errors of absolute Gaia $G$-band magnitudes and $G-G_{\rm RP}$ colors are typically very small (respective medians are 0.03\,mag and 0.04\,mag) and are mostly limited by measurement errors for the Gaia~DR2 bandpass zero points, but the spread of absolute $G$-band magnitudes within each color bins are larger, with standard deviations around the polynomial fits of 0.5\,mag for the youngest associations (10--15\,Myr) and 0.2\,mag otherwise. Those are most likely due to astrophysical phenomena such as circumstellar disks, differences in projection angles, rotational velocities, radii, and unresolved companions or background stars. Fitting polynomials to a binned version of the color-magnitude sequence is useful to prevent an over-fitting of the redder part of the color-magnitude diagram, because of the overwhelming fractional population of low-mass stars in these samples. The resulting color-magnitude sequences are displayed in Figure~1. We adopted the procedure described by \cite{2020ApJ...903...96G} to correct for extinction caused by interstellar dust and gas, which uses the STILISM 3D reddening maps \citep{2018AA...616A.132L}, and accounts for the wide Gaia photometric bandpasses in its reddening correction. This procedure only had a small but significant effect for the members of the Pleiades and Coma Berenices open clusters and the Sco-Cen OB association. As described in \cite{2020ApJ...903...96G}, the effect is generally a slight shift towards bluer Gaia~DR2 $G-G_{\rm RP}$ colors for OBA-type stars and a shift along empirical coeval sequences towards bluer $G-G_{\rm RP}$ colors and brighter absolute $G$-band magnitudes for K- and M-type stars.

\begin{figure}
 	\centering
 	\includegraphics[width=0.465\textwidth]{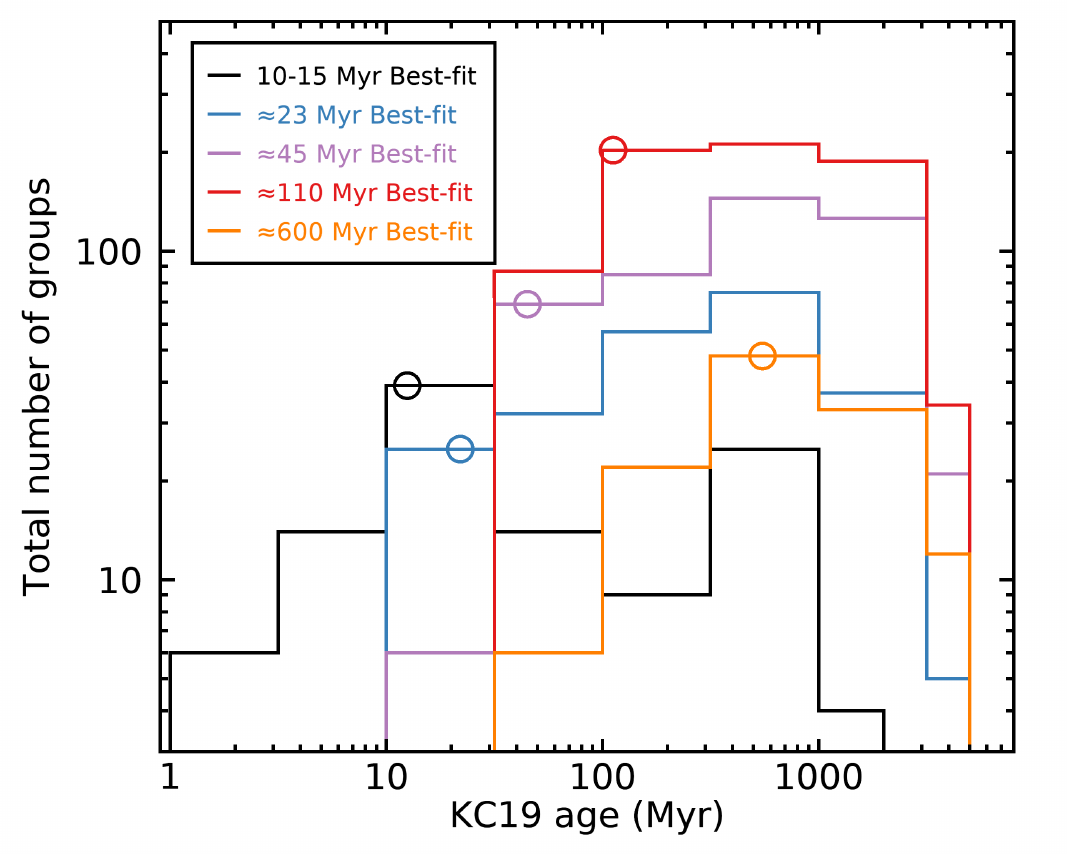}
 	\caption{Relative distribution of isochronal ages determined by KC19 for Theia groups, separated by categories of best-fitting empirical color-magnitude sequence. Color circles show the expected age of our empirical isochrones in each five categories; a perfect agreement would correspond to a single histogram bar lined up with the circle of each respective color. This figure shows that a significant number of Theia groups that we find are best matched by 10--110\,Myr empirical sequences were assigned much older ages in KC19. The number of such discrepant old ages is however smaller for the groups that we assign to younger empirical sequences.}
 	\label{fig:age_hist}
\end{figure}

Within each Theia group, we calculated the median value of the absolute vertical distance between every star and each one of the empirical age sequences in the color-magnitude diagram. We then selected the case with the smallest median absolute distance as the `best-fitting' age. We note here that this method is not intended to provide accurate age measurements because of the coarse selection between five models only. The empirical sequence of 600\,Myr-old stars in a Gaia absolute $G$ versus $G-G_{\rm RP}$ color-magnitude diagram is similar to that of older unrelated field stars, and we therefore expect that Theia groups older than 600\,Myr will be assigned a best-fitting age of 600\,Myr with this method. The advantage of this method is that the results should not suffer from biases of models underlying classical isochrones, and they will be less susceptible to outliers. In Figure~\ref{fig:age_hist}, we show our best-fitting ages compared with those of KC19. 

We also observed that several of the Theia groups were assigned average $XYZUVW$ values that were sensitive to a few visual outliers; to facilitate our investigation, we therefore also calculated the median $XYZUVW$ value for each Theia group. The next step in our analysis consisted in pre-selecting only Theia groups that have both a best-matching empirical isochrone and median $UVW$ space velocities consistent to those of known young association listed in \cite{2018ApJ...856...23G}, and visualizing them in $XYZ$ coordinates with Partiview. We initially visualized all groups Theia with $UVW$ velocities matching within 10\,\kms\ of a known association with an age not more than one empirical sequence away from that of the group. Any Theia groups that seemed possibly connected to a known young association, therefore potentially forming a spatial extension to the group, were marked as interesting and we also visualized the color-magnitude diagram of their members to ensure that our empirical sequence selection described above yielded a sensible age. In addition to this, we visualized the $UVW$ positions of the members of interesting Theia groups using Partiview, to ensure that the distribution was not too scattered, and consistent with that of the potentially related known young association.

\section{CANDIDATE SYSTEMS OF DISSOLVING OPEN CLUSTERS}\label{sec:results}

In Table~\ref{tab:list}, we present the list of moving groups and Theia structures that we determined may correspond to spatial extensions of larger young associations or open clusters, with their median $UVW$ space velocities and estimated ages. In Figure~\ref{fig:xy}, we compare the $XY$ Galactic coordinates of these potentially related groups, from which we performed a kinematic-based pre-selection using the relevant BANYAN~$\Sigma$ $UVW$ model \citep{2018ApJ...856...23G} and the best case scenario $UVW$ separation between the members of a Theia group and other potentially related young associations. Any Theia star separated by more than 5\,\kms\ in $UVW$ space (about three times the standard deviation of well-behaved young associations) was therefore omitted, to reduce the non-negligible contamination observed in several Theia groups. We show the common  space velocities of plausibly related groups in Figure~\ref{fig:uv}, this time including even the kinematic outliers of Theia groups. For example, the members of Theia~301 separated by more than 5\,\kms\ from the center of the BANYAN~$\Sigma$ $UVW$ model of the Pleiades were ommitted from Figure~\ref{fig:xy}, but are shown in Figure~\ref{fig:uv}. A detailed description of the kinematics and sample construction of the Theia groups is provided in KC19, but we note that the $XYZ$ Galactic coordinates and $UVW$ space velocities for the Theia groups discussed here are based on Gaia~DR2 data alone and have measurement errors in the range 0.1--2.0\,pc and 0.2--1.0\,\kms, respectively. We observe a similar range of measurement errors in the case of other associations described here as they are also mostly based on Gaia~DR2 data. Furthermore, only Theia stars with radial velocity measurements in Gaia~DR2 have $UVW$ errors, and as a consequence only $\approx$\,18\%\ of the stars in relevant Theia groups are represented in Figure~\ref{fig:uv}.

We have tentatively named each of the potentially related ensembles of young associations as `systems'. We have chosen this particular terminology because we consider that these hypotheses still need to be confirmed through additional observations, but we would encourage use of the terminology of star-formation complexes, open clusters, tidal tails or coronae where relevant (e.g., \citealt{2021AA...645A..84M}) if future observations confirm them. In the remainder of this section, we discuss six candidate `systems' of related groups that we have identified.

\subsection{The Pleiades system}\label{sec:ple}

The AB~Doradus moving group has long been hypothesized to be related to the Pleiades open cluster (e.g., \citealt{2005ApJ...628L..69L}), but the limited spatial distribution of its known members made this possibility less compelling, although it could be expected that the present-day spatial extent of the AB~Doradus moving group may currently be observationally biased. The inclusion of Theia~301 and Theia~369 members that have space velocities consistent with the Pleiades (i.e., the vast majority of them, see Figure~\ref{fig:abdmg_uv}) with the spatial distribution of Pleiades members as shown in Figure~\ref{fig:abdmg_xy} makes this possibility more compelling. Theia~369 contains many known members of the Pleiades, but extends to further distances. Theia~301 may consititute the trailing tidal tail of the Pleiades, and the AB~Doradus moving group might in fact be a well-explored section of this tidal tail, due to its proximity and historical scrutiny. If these structures in fact constitute a trailing tidal tail of the Pleiades, we might expect to find a leading tail at further distances behind the Pleiades, which might not have yet been uncovered due to sparsely available radial velocities in Gaia~DR2, or simply due to the difficulty in sampling the region of space located behind the Pleiades. These considerations could explain why \cite{2021AA...645A..84M} did not recover tidal tails to the Pleiades.

Our empirical color-magnitude analysis suggests that Theia~234 has a younger age of $\sim$\,45\,Myr, in principle inconsistent with that of the Pleiades, but given its spatial position suggestive of a possible extension of Theia~301 and its similar space velocities, we consider that it should be studied further to determine whether it is also associated with the Pleiades.

We note that another KC19 group, Theia~368, appears to have an age ($\sim$\,110\,Myr) and space velocities consistent with the Pleiades, but it is located near the much younger Sco-Cen OB association with a significant spatial gap between it and all other structures of the Pleiades system proposed here. It may be foreseeable that more members are missing between the AB~Doradus moving group and Theia~368, but verifying this will likely be challenging because of the cross-contamination that may be expected from Sco-Cen. We also note that the combination of Theia~301 and Theia~368 would appear to constitute an unexpected shape if they collectively formed a structure of tidal tails around the Pleiades (Figure~\ref{fig:abdmg_xy}).

\startlongtable
\begin{deluxetable*}{lccl}
\tablecolumns{4}
\tablecaption{Systems of of plausibly related groups.\label{tab:list}}
\tablehead{\colhead{Group} & \colhead{Age (Myr)} & \colhead{$UVW$ (\kms)} & \colhead{References}}
\startdata
\textbf{Pleiades} & $112 \pm 5$ & ($-6.7$,$-28.0$,$-14.0$) & --,1,2\\
AB~Doradus moving group & $133^{+15}_{-20}$ & ($-7.2$,$-27.6$,$-14.2$) & 3,4,2\\
Theia~301 & $\approx$\,110 & ($-7.2$,$-26.6$,$-13.1$) & 5,19,19\\
Theia~369 & $\approx$\,110 & ($-6.9$,$-28.4$,$-14.2$) & 5,19,19\\
Theia~368 & $\approx$\,110 & ($-1.4$,$-28.7$,$-13.8$) & 5,19,19\\
Theia~234 & $\approx$\,45 & ($-10.9$,$-26.5$,$-12.6$) & 5,19,19\\
\tableline
\textbf{IC~2602} & $46_{-5}^{+6}$ & ($-8.2$,$-20.6$,$-0.6$) & --,6,2\\
IC~2602 corona & $\cdots$ & $\cdots$ & 11,--,--\\
Tucana-Horologium & $45 \pm 4$ & ($-9.8$,$-20.9$,$-1.0$) & 7,8,2\\
Theia~92 (partial) & $\approx$\,23 & ($-10.3$,$-21.9$,$-3.0$) & 5,19,19\\
\tableline
\textbf{IC~2391} & $50 \pm 5$ & ($-23.0$,$-14.9$,$-5.5$) & --,9,2\\
IC~2391 corona & $\cdots$ & $\cdots$ & 11,--,--\\
Argus & 40--50 & ($-22.5$,$-14.6$,$-5.0$) & 10,12,12\\
Theia~114 & $\approx$\,45 & ($-23.8$,$-14.7$,$-5.6$) & 5,19,19\\
Theia~115 & $\approx$\,45 & ($-23.3$,$-14.7$,$-6.0$) & 5,19,19\\
\tableline
\textbf{Octans} & 30--40 & ($-13.7$,$-3.3$,$-10.1$) & 14,15,2\\
Octans-Near & 30--100 & ($-13.1$,$-3.7$,$-10.7$) & 16,16,19\\
Theia~94 & $\approx$\,45 & ($-11.5$,$-3.0$,$-9.1$) & 5,19,19\\
Theia~95 (partial) & $\approx$\,45 & ($-17.4$,$-5.6$,$-10.9$) & 5,19,19\\
Theia~44 & $\approx$\,23 & ($-12.8$,$-6.8$,$-9.1$) & 5,19,19\\
Platais~5 & $\approx$\,60 & ($-20.8$,$-5.7$,$-9.3$) & 13,13,19\\
\tableline
\textbf{Platais~8} & $\approx$\,60 & ($-11.0$,$-22.9$,$-3.6$) & 13,13,2\\
Carina & $45_{-7}^{+11}$ & ($-10.7$,$-21.9$,$-5.5$) & 14,8,2\\
Columba & $42_{-4}^{+6}$ & ($-11.9$,$-21.3$,$-5.7$) & 14,8,2\\
Theia~92 (partial) & $\approx$\,23 & ($-10.3$,$-21.9$,$-3.0$) & 5,19,19\\
Theia~113 & $\approx$\,45 & ($-11.5$,$-21.8$,$-3.7$) & 5,19,19\\
Theia~208 & $\approx$\,45 & ($-14.5$,$-22.2$,$-5.9$) & 5,19,19\\
\tableline
\textbf{32~Ori} & $22_{-3}^{+4}$ & ($-12.8$,$-18.8$,$-9.0$) & 17,8,2\\
$\beta$ Pictoris moving group & $24 \pm 3$ & ($-10.9$,$-16.0$,$-9.0$) & 18,8,2\\
Theia~62 & $\approx$\,22 & ($-8.1$,$-15.9$,$-7.6$) & 5,19,19\\
Theia~65 & $\approx$\,22 & ($-11.8$,$-18.5$,$-8.8$) & 5,19,19\\
\enddata
\tablecomments{References are cited for the discovery, age, and $UVW$ velocities, respectively. Ages cited here are directly taken from the relevant references, and thus the more poorly studied young associations for which only approximate ages have been estimated in the literature do not have measurement errors associated with them.}
\tablerefs{(1)~\citealt{2015ApJ...813..108D}; (2)~\citealt{2018ApJ...856...23G}; (3)~\citealt{2004ApJ...613L..65Z}; (4)~\citealt{2018ApJ...861L..13G}; (5)~\citealt{2019AJ....158..122K}; (6)~\citealt{2010MNRAS.409.1002D}; (7)~\citealt{2000AJ....120.1410T} and \citealt{2001ASPC..244..122Z}; (8)~\citealt{2015MNRAS.454..593B}; (9)~\citealt{2004ApJ...614..386B}; (10)~\citealt{2000MNRAS.317..289M}; (11)~\citealt{2021AA...645A..84M}; (12)~\citealt{2019ApJ...870...27Z}; (13)~\citealt{1998AJ....116.2423P}; (14)~\citealt{2008hsf2.book..757T}; (15)~\citealt{2015MNRAS.447.1267M}; (16)~\citealt{2013ApJ...778....5Z}; (17)~\citealt{2007IAUS..237..442M}; (18)~\citealt{2001ApJ...562L..87Z}; (19)~This paper.}
\end{deluxetable*}

\subsection{The IC~2602 system}\label{sec:ic2602}

While the Columba, Carina and Tucana-Horologium associations have previously been suggested to be potentially related to one another \citep{2008hsf2.book..757T}, no previous studies seem to have provided a detailed discussion of the surprising similarity in space velocities and ages between the Columba, Carina, Tucana-Horologium associations and the Platais~8 and IC~2602 open clusters.

We propose that the Tucana-Horologium association may be part of a tidal tail associated with IC~2602, and that both of these are dissociated from Columba, Carina and Platais~8. The distinct $X$ component of their Galactic coordinates and the $W$ component of their space velocities (see Figures~\ref{fig:col_uv} and \ref{fig:tha_uv}) seem to warrant such a separation between what we tentatively call the IC~2602 system and the Platais~8 system (described below). For this reason, we choose to display $U$ and $W$ components of space velocities in Figures~\ref{fig:col_uv} and \ref{fig:tha_uv}, rather than $U$ and $V$. We have also selected a 2.5\,\kms\ kinematic cut-off to choose which members of Theia groups are displayed in which panel of Figure~\ref{fig:xy} to reduce cross-contamination between the IC~2602 and Platais~8 systems. If both the IC~2602 and Platais~8 systems are confirmed to be respectively physical, it will be interesting to investigate whether their full spatial and kinematic distributions are consistent with a past or ongoing disruptive interaction between IC~2602 and Platais~8.

We have only identified one KC19 structure (parts of Theia~92) that seems to be related to IC~2602 and the Tucana-Horologium association, and we find that it mostly corresponds to the IC~2602 open cluster itself. While we have not identified additional KC19 groups that bridge the physical gap between these two groups, we suggest that this may only be due to projection effects. It will be useful to investigate whether additional co-moving stars can be uncovered that may fill the gap between the inner edge of the \cite{2021AA...645A..84M} IC~2602 corona and known Tucana-Horologium members (see Figure~\ref{fig:tha_xy}).

\subsection{The Platais~8 system}\label{sec:pl8}

The Platais~8 open cluster was discovered by \cite{1998AJ....116.2423P} and has received relatively little attention since. It was included in the sample of \cite{2018AA...615A..12Y}, but not those of \cite{2021AA...645A..84M} and \cite{2021ApJS..253...38L}. Theia~92 seems to include the core of Platais~8 and a possible trailing tidal tail behind this core (see Figure~\ref{fig:col_xy}). Remarkably, another KC19 group (Theia~113) fills the spatial gap between Theia~92 and members of Columba and Carina, and has both an age and space velocities consistent with them. Theia~208 also shows consistent properties and may constitute the tip of the leading tidal tail of Platais~8, which would be formed by the combination of Theia~113, Columba, Carina, and Theia~208.

\subsection{The IC~2391 system}\label{sec:ic2391}

The Argus association described by \cite{2000MNRAS.317..289M} was hypothesized to be related to the IC~2391 cluster because of its similar age and space velocities. The existence of the Argus association was recently challenged due to the seemingly non-coeval nature of its members when their color-magnitude sequence was studied by \cite{2015MNRAS.454..593B}, but a further study by \cite{2019ApJ...870...27Z} demonstrated that this was likely caused by a high level of contamination in previous membership lists of Argus, and they assembled a set of well-behaved co-moving and seemingly co-eval stars using Gaia~DR2 data.

Here, we find that Theia~114 is consistent with containing most of the core of IC~2391, and the leading part of the IC~2391 tidal tails (or `corona') discovered by \cite{2021AA...645A..84M} seem to extend all the way to the furthest known members of the Argus association defined by \citeauthor{2019ApJ...870...27Z} (\citeyear{2019ApJ...870...27Z}; see Figure~\ref{fig:arg_xy}). We find that Theia~115 also seems to have a consistent age and kinematics, but its distribution of space velocities (see Figure~\ref{fig:arg_uv}) suggest that it may be highly contaminated by unrelated field stars, as was the case with the earlier instantiations of the Argus association.

\subsection{The Octans system}\label{sec:octans}

\cite{2013ApJ...778....5Z} uncovered a nearby set of stars co-moving and seemingly co-eval with the known Octans association \citep{2008hsf2.book..757T}, which they proposed as a new association named Octans-Near. Here, we find that a large fraction of the KC19 group Theia~95, as well as the smaller group Theia~94, may also be related to Octans and Octans-Near, based on their similar ages, space velocities (Figure~\ref{fig:oct_uv}) and the fact that they form a spatial extension to Octans and Octans-Near in the $XY$ plane (see Figure~\ref{fig:oct_xy}). The poorly studied Platais~5 open cluster candidate seems to be consistent with the age and kinematics of these groups, and falls remarkably squarely within the spatial distribution of Theia~95. The $U$ component of the space velocity of Platais~5 appear to be discrepant (by $\approx$7.5\,\kms), but it is currently based on the only two known members with full kinematics (HIP~29306 and HIP~29941) -- here we listed the average velocities of these two members as our best estimate of the Platais~5 kinematics in Table~\ref{tab:list}.  While only six B- and A-type members of Platais~5 are well-documented, it is likely that they constitute the tip of the iceberg of the Platais~5 open cluster in terms of its members.

We find that Theia~44 has consistent kinematics with our proposed wider Octans system, but its well-concentrated spatial distribution does not form a compelling extension to Octans or Octans-Near (although three known members of Octans-Near are possibly more related to Theia~44 than other members of Octans-Near, see Figure~\ref{fig:oct_xy}). Our empirical age estimation for Theia~44 is also slightly younger than other components described here ($\approx$\,22\,Myr versus 40--60\,Myr), but we still consider future studies of Theia~44 to be warranted given its similarities to other groups in the Octans system.

\subsection{The 32~Ori system}\label{sec:thor}

\begin{figure*}[p]
	\centering
	\subfigure[Pleiades system]{\includegraphics[width=0.45\textwidth]{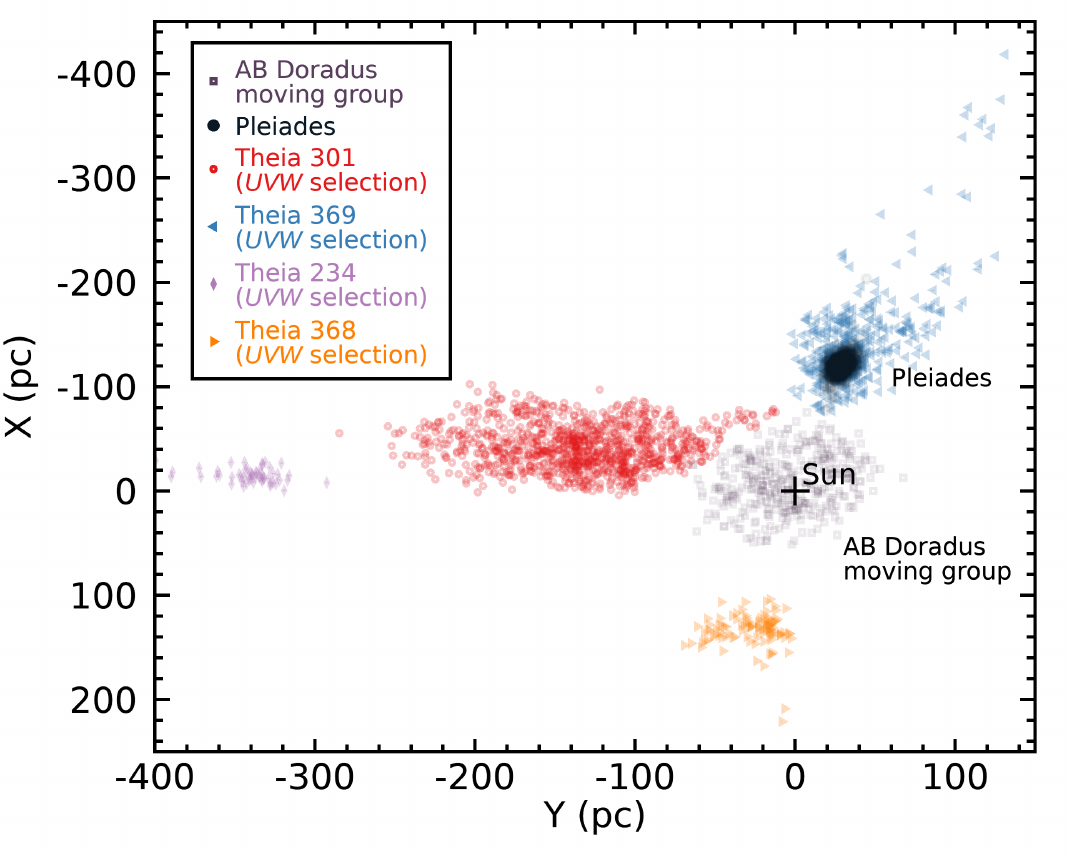}\label{fig:abdmg_xy}}
	\subfigure[Platais~8 system]{\includegraphics[width=0.45\textwidth]{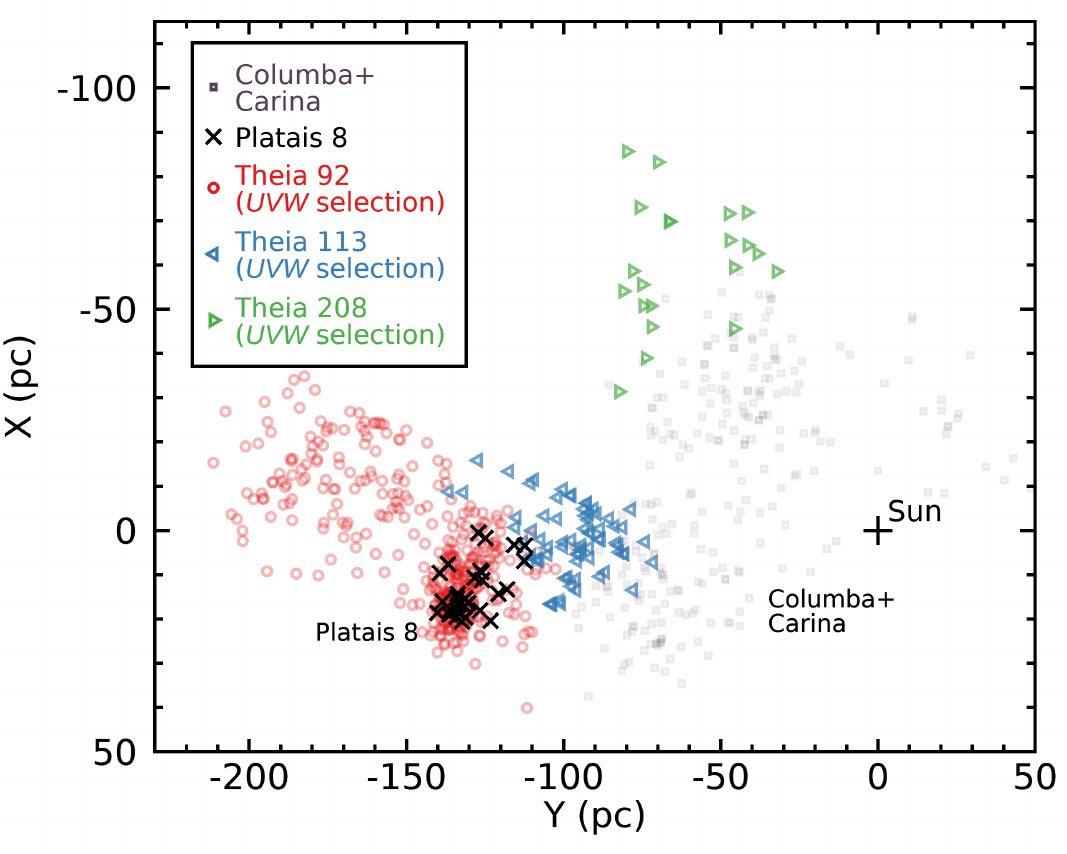}\label{fig:col_xy}}
	\subfigure[IC~2602 system]{\includegraphics[width=0.45\textwidth]{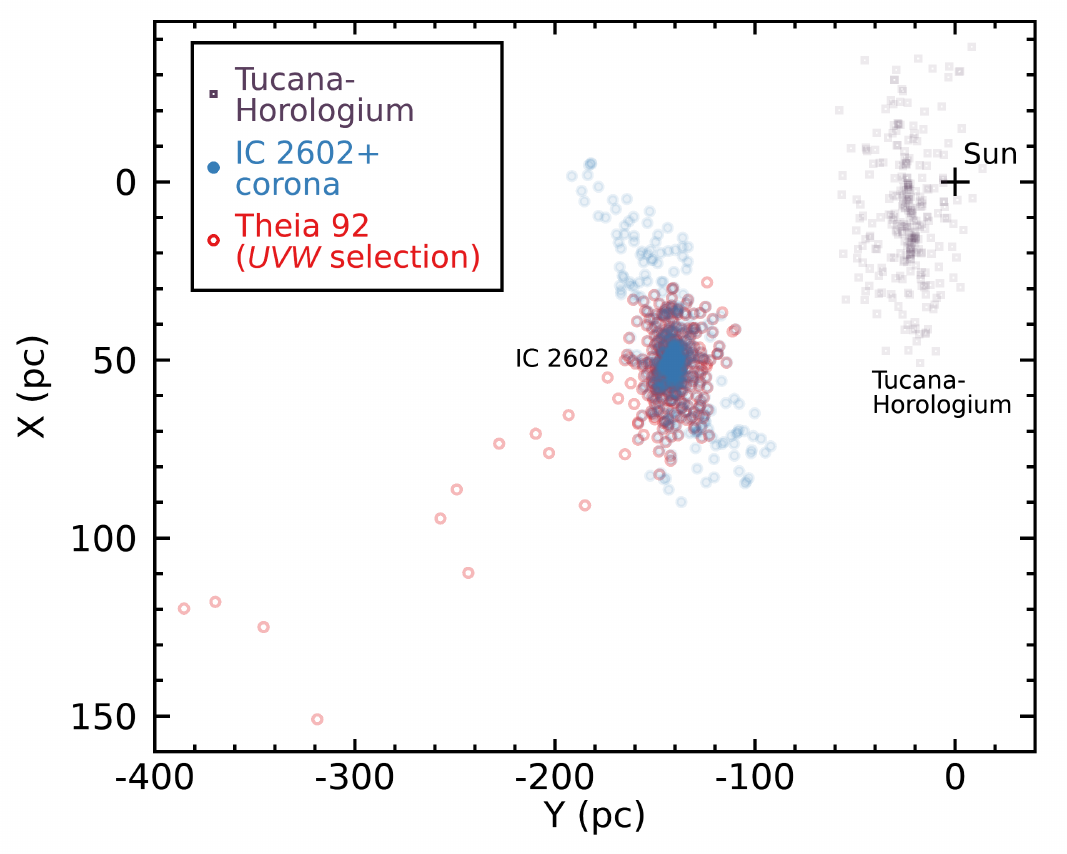}\label{fig:tha_xy}}
	\subfigure[IC~2391 system]{\includegraphics[width=0.45\textwidth]{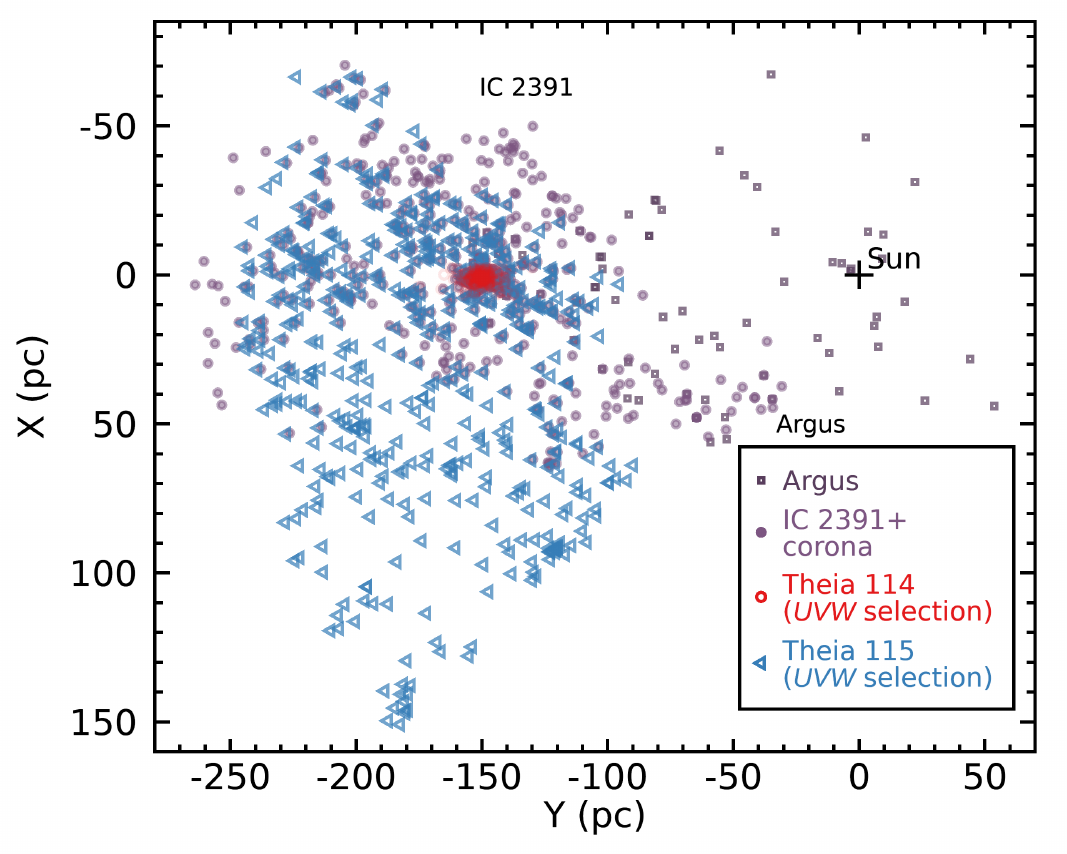}\label{fig:arg_xy}}
	\subfigure[Octans system]{\includegraphics[width=0.45\textwidth]{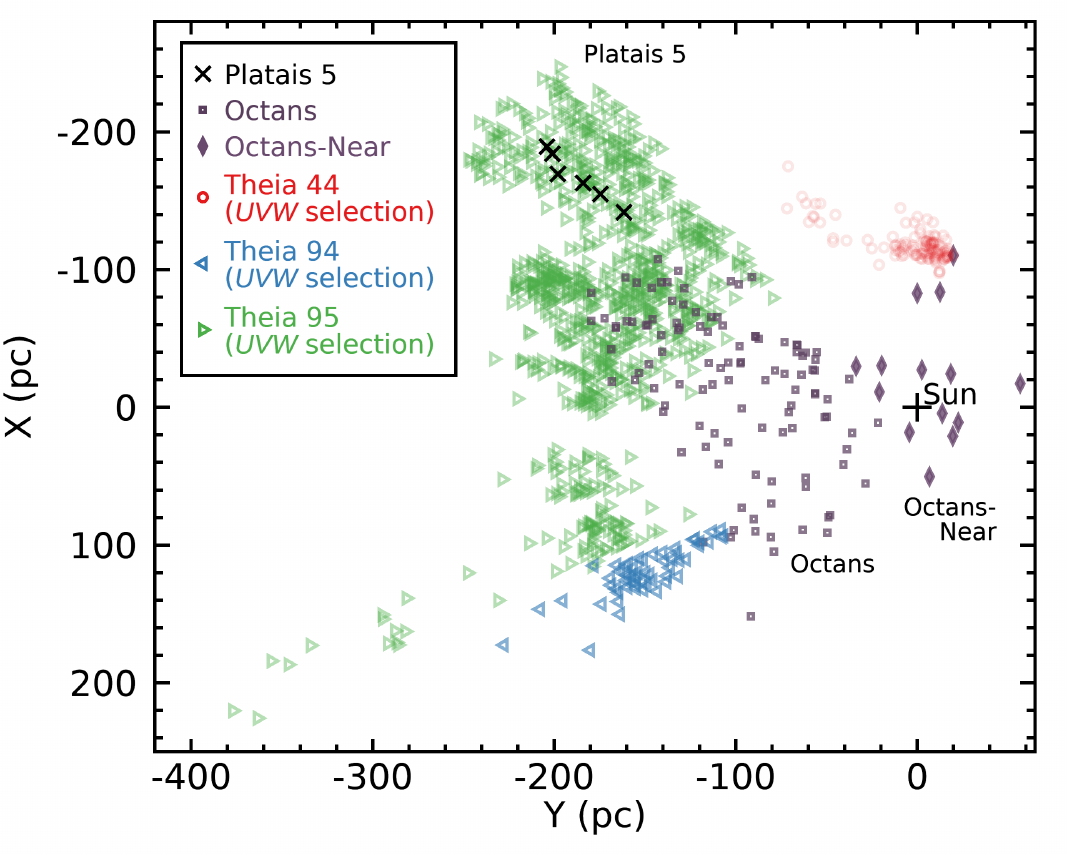}\label{fig:oct_xy}}
	\subfigure[32~Ori system]{\includegraphics[width=0.45\textwidth]{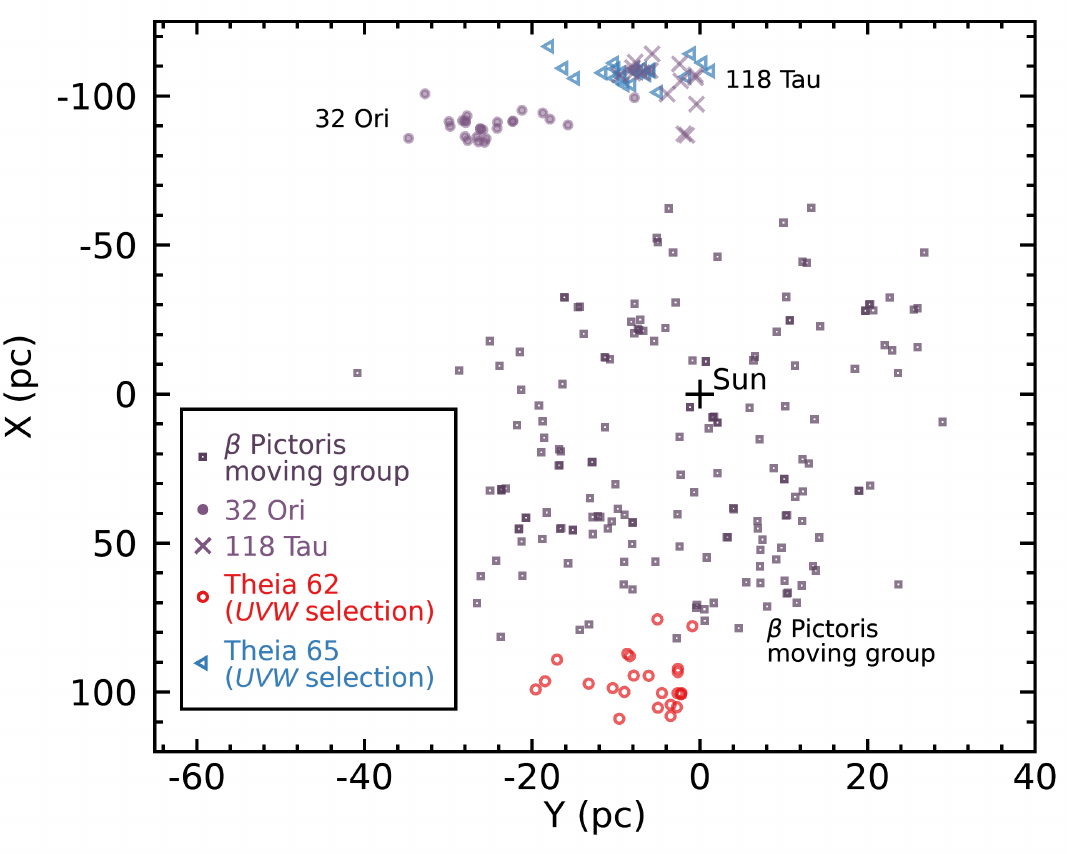}\label{fig:bpmg_xy}}
	\caption{$XY$ Galactic coordinates of plausibly related systems of open clusters, moving groups and young associations. Typical measurement errors are 0.1--2.0\,pc. See Section~\ref{sec:results} for more details.}
	\label{fig:xy}
\end{figure*}

\begin{figure*}[p]
	\centering
	\subfigure[Pleiades system]{\includegraphics[width=0.44\textwidth]{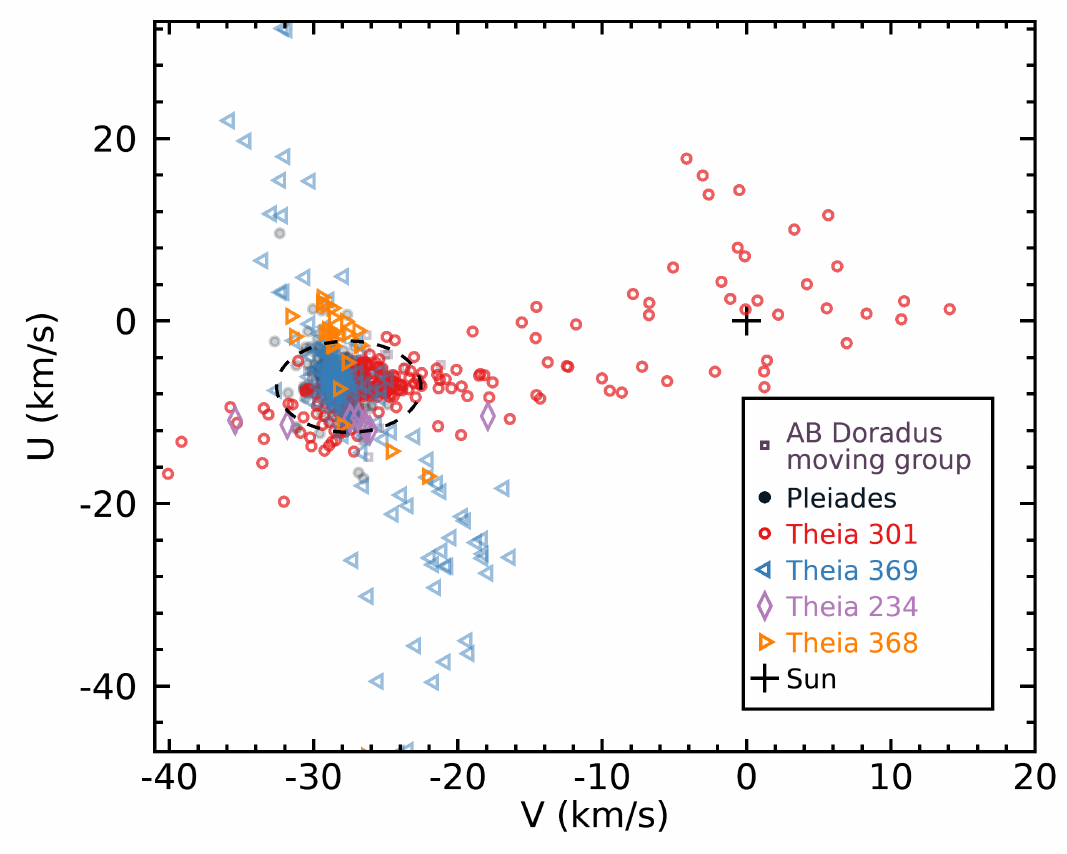}\label{fig:abdmg_uv}}
	\subfigure[Platais~8 system]{\includegraphics[width=0.44\textwidth]{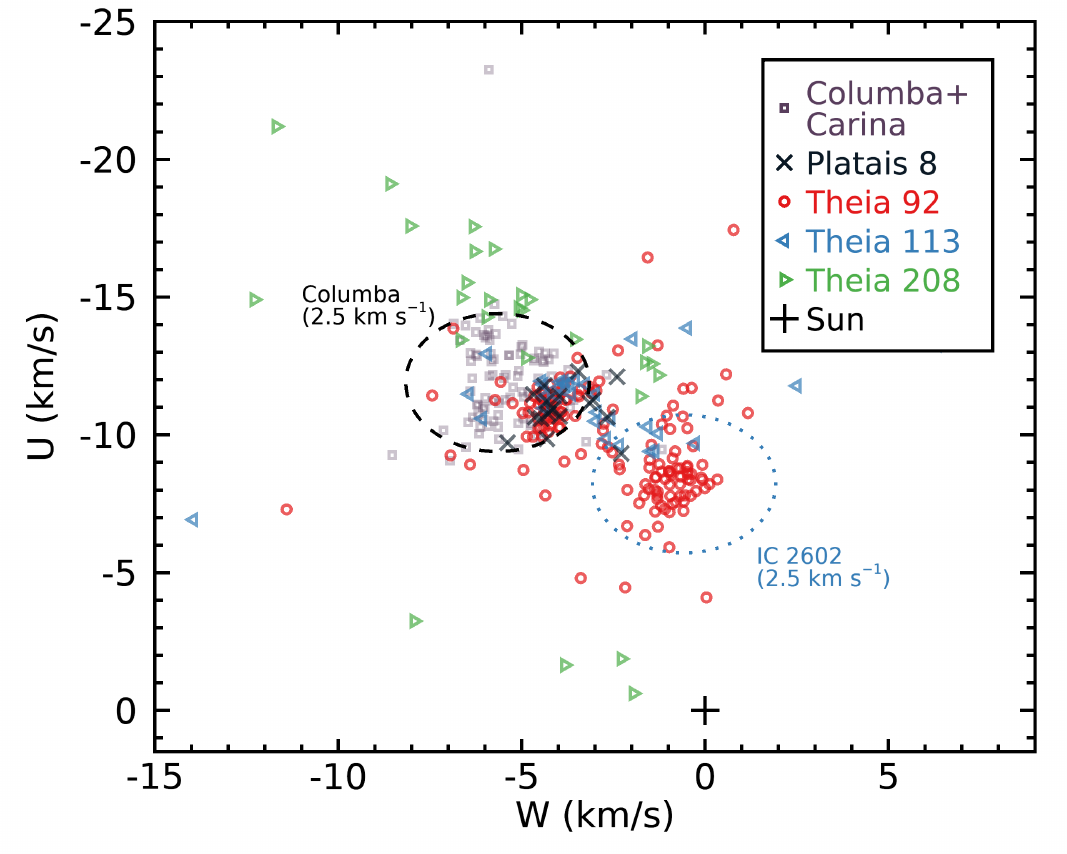}\label{fig:col_uv}}
	\subfigure[IC~2602 system]{\includegraphics[width=0.44\textwidth]{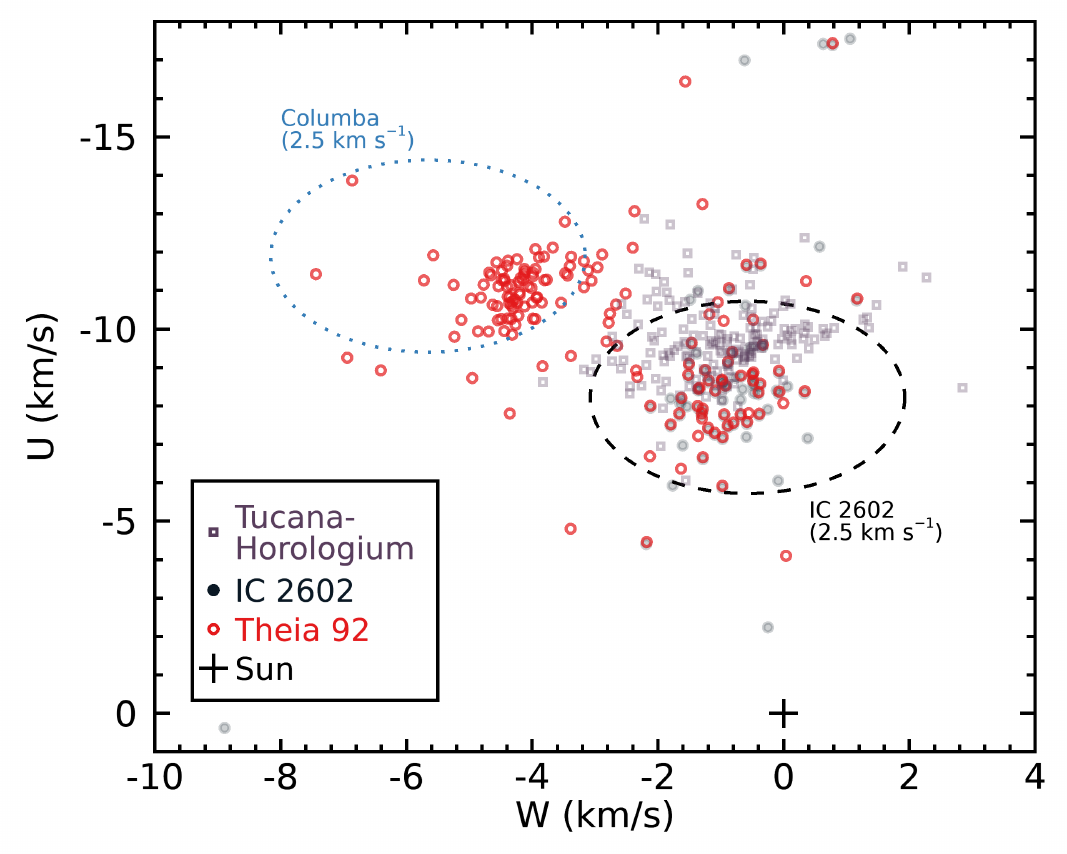}\label{fig:tha_uv}}
	\subfigure[IC~2391 system]{\includegraphics[width=0.44\textwidth]{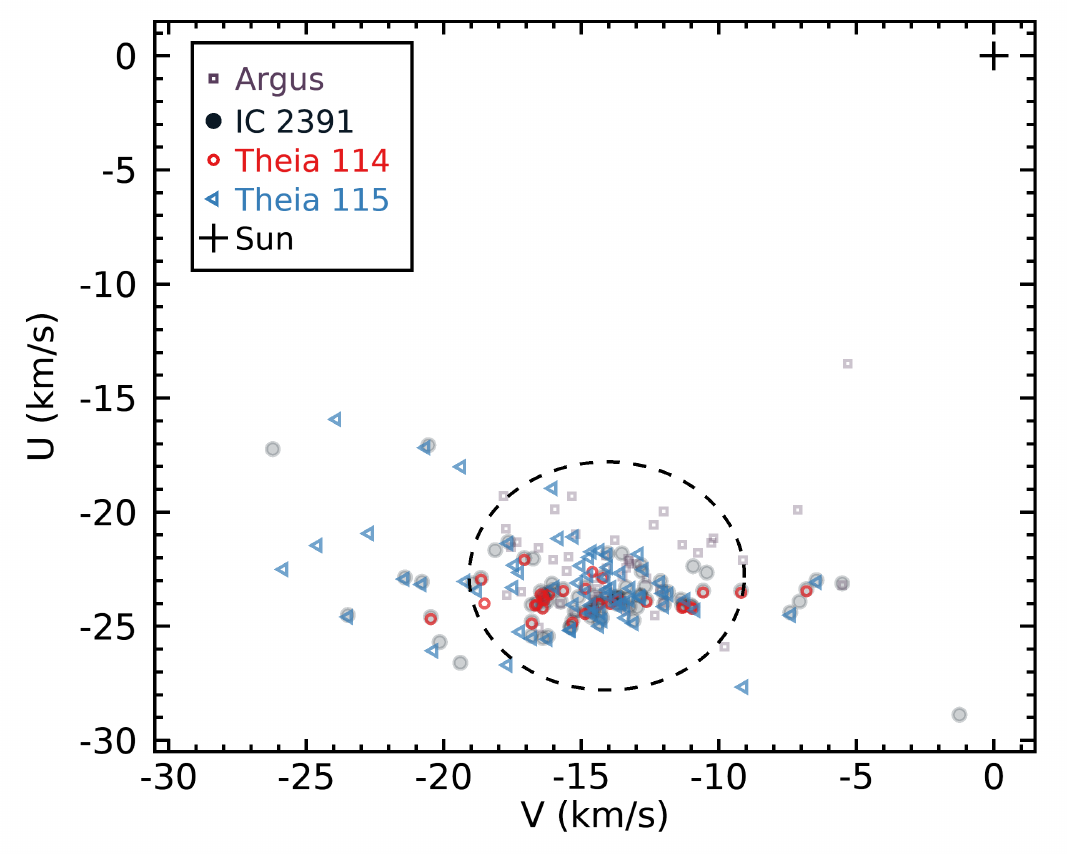}\label{fig:arg_uv}}
	\subfigure[Octans system]{\includegraphics[width=0.44\textwidth]{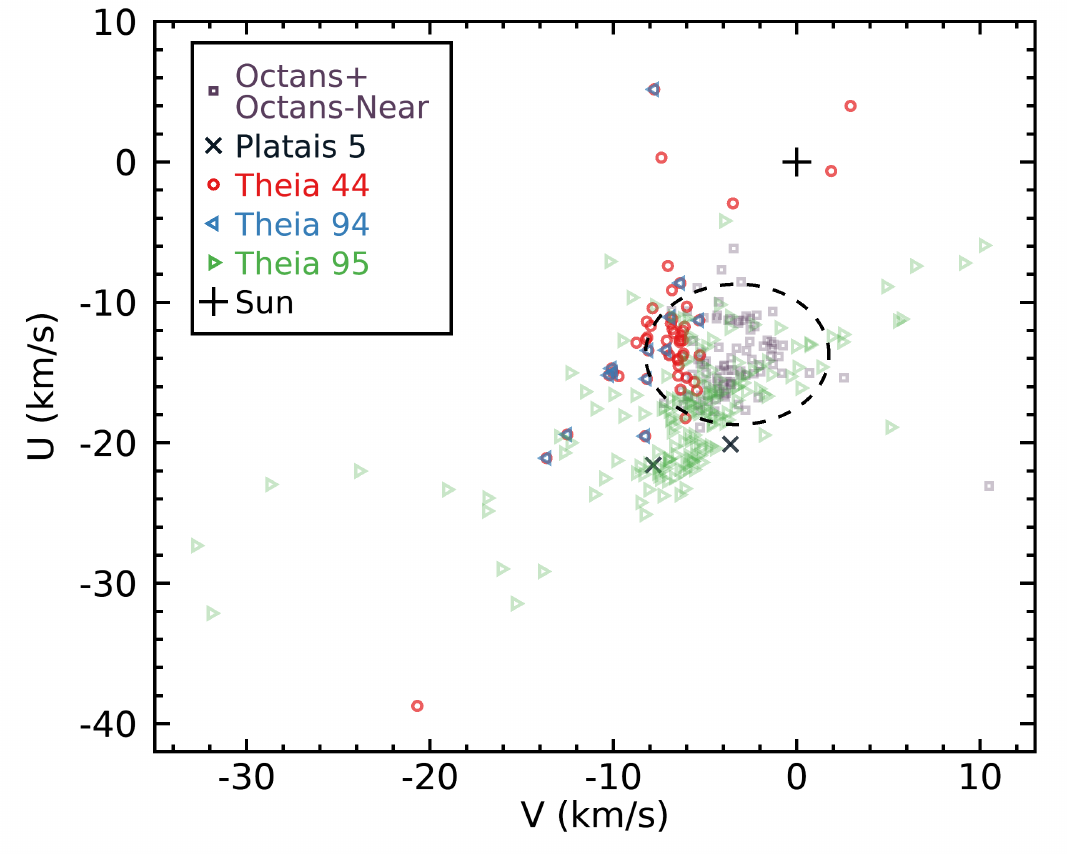}\label{fig:oct_uv}}
	\subfigure[32~Ori system]{\includegraphics[width=0.44\textwidth]{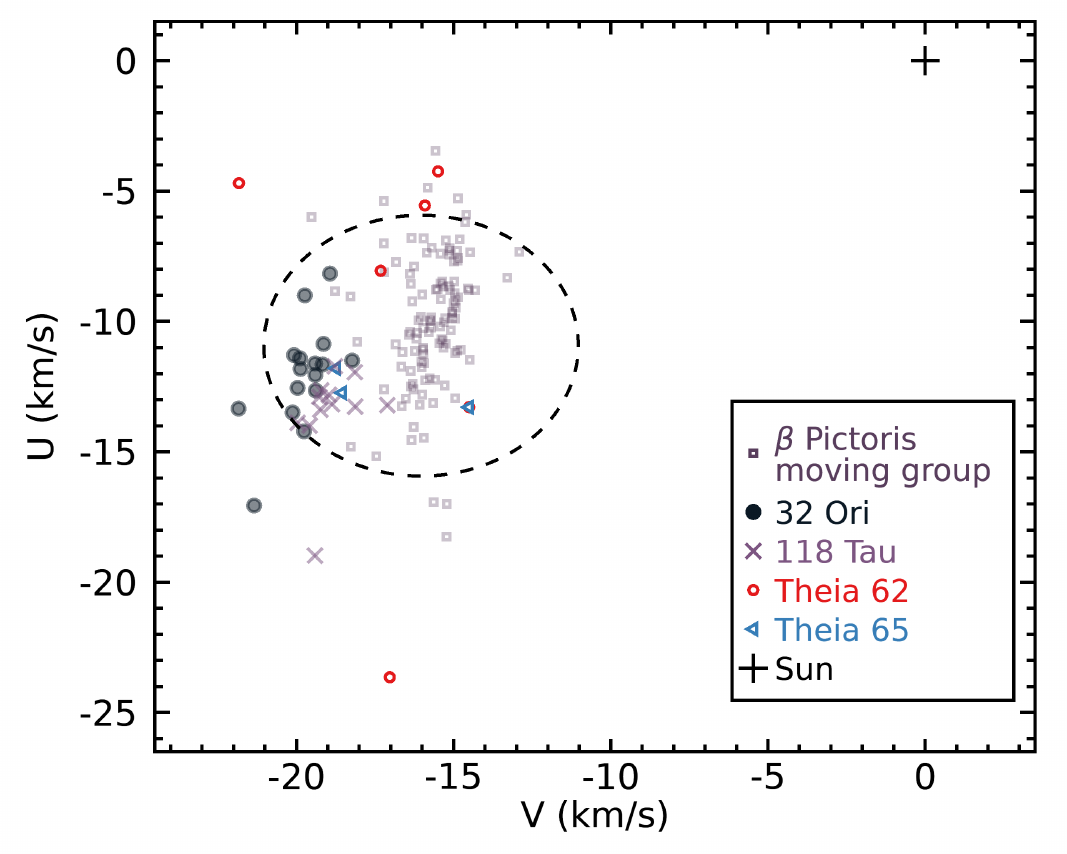}\label{fig:bpmg_uv}}
	\caption{$UVW$ space velocities of plausibly related systems of open clusters, moving groups and young associations. Unless otherwise noted, the dashed circles represent the 5\,\kms\ limits around the center of the relevant BANYAN~$\Sigma$ $UVW$ model used to filter out kinematic outliers from Theia groups. Typical measurement errors are 0.2--1.0\,\kms. See Section~\ref{sec:results} for more details.}
	\label{fig:uv}
\end{figure*}

The $\beta$~Pictoris moving group \citep{2001ApJ...562L..87Z} has been the focus of intensive scrutiny. Its very close proximity and young age make it a particularly valuable laboratory to search for exoplanets by direct-imaging, illustrated by the discoveries of $\beta$~Pictoris~b \citep{2010Sci...329...57L} and AU~Mic~b \citep{2020Natur.582..497P} and isolated substellar objects of very low masses \citep{2013ApJ...777L..20L}. While we have identified a single KC19 group, Theia~65, that seems physically overlapping with members of the $\beta$~Pictoris moving group (see Figure~\ref{fig:bpmg_xy}), we suggest that the 32~Ori association described by \cite{2007IAUS..237..442M} may be related to the $\beta$~Pictoris moving group due to its similar age and kinematics, although their $UVW$ space velocities differ by about 3.4\,\kms. The current gap between the spatial distributions of these two groups is located on the edge of where clustering methods in direct observable space become inefficient ($\approx$\,80\,pc), and it is plausible that a number of co-moving stars remain to be identified between 32~Ori and the $\beta$~Pictoris moving group. If the latter is indeed a leading tidal tail of the 32~Ori association, it will also become interesting to assign further scrutiny to stars in the background of 32~Ori which may include a trailing tidal tail.

We note that Theia~65 has an age and kinematics consistent with the $\beta$~Pictoris moving group and the 32~Ori association, but a similar gap exists between it and 32~Ori (see Figure~\ref{fig:bpmg_xy}), and Theia~65 seems spatially more similar to the somewhat younger $\approx$10\,Myr-old 118~Tau association (see \citealt{2018ApJ...856...23G} for more details). Perhaps the age of 118~Tau was previously under-estimated, but it is also plausible that neither 118~Tau or Theia~65 are related to the $\beta$~Pictoris moving group or the 32~Ori association.

\section{CONCLUSIONS}\label{sec:conclusion}

We presented tentative evidence based on kinematics and color-magnitude diagrams to suggest that several nearby moving groups may be spatially much more extended than previously thought, and that some of them might consist of tidal dissipation tails co-eval and co-moving with the cores of currently known open clusters. Further observations such as radial velocity and lithium-based age-dating will be required to corroborate this. New radial velocities in Gaia~DR3 in particular will be helpful to refine the kinematics of the groups discussed here. Data from the TESS mission \citep{2014SPIE.9143E..20R} will be helpful to age-date these populations with color-rotation period diagrams, and the eROSITA mission \citep{Cappelluti:2010uy} will also provide valuable X-ray luminosities that will serve to further assess the ages of these stellar populations. In addition to these future observations, clustering methods that are specifically designed to uncover nearby sparse moving groups with partially missing kinematics will be key to complete the spatial mapping of these potential large co-moving structures. The Platais~5 and Platais~8 open clusters, which we find are potentially related to known young moving groups near the Sun, are also currently poorly characterized and will require dedicated follow-up studies.

\acknowledgments

We would like to thank our anonymous reviewer for their thoughtful comments that improved the quality of this Letter. This research made use of the SIMBAD database and VizieR catalog access tool, operated at the Centre de Donn\'ees astronomiques de Strasbourg, France \citep{2000AAS..143...23O}. This work presents results from the European Space Agency (ESA) space mission Gaia. Gaia data are being processed by the Gaia Data Processing and Analysis Consortium (DPAC). Funding for the DPAC is provided by national institutions, in particular the institutions participating in the Gaia MultiLateral Agreement (MLA). The Gaia mission website is https://www.cosmos.esa.int/gaia. The Gaia archive website is https://archives.esac.esa.int/gaia.

\bibliographystyle{apj}

\end{document}